\documentclass[reprint,
superscriptaddress,
nobalancelastpage,
 amsmath,amssymb,
 notitlepage,
10pt, 
tightenlines
]{revtex4-2}

\usepackage[utf8]{inputenc}
\usepackage[caption=false]{subfig}
\usepackage{graphicx}
\usepackage{dcolumn}
\usepackage{bm}
\usepackage{xcolor}
\usepackage{verbatim}
\usepackage{tabularx}
\usepackage{bbold}
\usepackage{float}
\usepackage{bm}
\usepackage{bbm}
\usepackage[colorlinks=True, citecolor=blue, urlcolor=blue]{hyperref}
\usepackage{booktabs}
\usepackage{natbib}

\usepackage{braket}

\definecolor{crimson}{rgb}{.8, 0, 0}

\newtheorem{theorem}{Theorem}

\begin{document}
\title{Circumventing superexponential runtimes for hard instances of quantum adiabatic optimization}

\author{Benjamin~F.~Schiffer}
\email[Corresponding author: ]{Benjamin.Schiffer@mpq.mpg.de}
\affiliation{Max-Planck-Institut f\"ur Quantenoptik, Hans-Kopfermann-Str.~1, D-85748 Garching, Germany}%
\affiliation{Munich Center for Quantum Science and Technology (MCQST), Schellingstr.~4, D-80799 Munich, Germany}%
\author{Dominik S.~Wild}%
\affiliation{Max-Planck-Institut f\"ur Quantenoptik, Hans-Kopfermann-Str.~1, D-85748 Garching, Germany}%
\affiliation{Munich Center for Quantum Science and Technology (MCQST), Schellingstr.~4, D-80799 Munich, Germany}%
\author{Nishad Maskara}%
\affiliation{Department of Physics, Harvard University, Cambridge, MA 02138, USA}%
\author{Madelyn Cain}%
\affiliation{Department of Physics, Harvard University, Cambridge, MA 02138, USA}%
\author{Mikhail D.~Lukin}%
\affiliation{Department of Physics, Harvard University, Cambridge, MA 02138, USA}%
\author{Rhine Samajdar}%
\affiliation{Department of Physics, Princeton University, Princeton, NJ 08544, USA}%
\affiliation{Princeton Center for Theoretical Science, Princeton University, Princeton, NJ 08544, USA}%

\date{\today}

\begin{abstract}
Classical optimization problems can be solved by adiabatically preparing the ground state of a quantum Hamiltonian that encodes the problem. The performance of this approach is determined by the smallest gap encountered during the evolution. Here, we consider the maximum independent set problem, which can be efficiently encoded in the Hamiltonian describing a Rydberg atom array. We present a general construction of instances of the problem for which the minimum gap decays superexponentially with system size, implying a superexponentially large time to solution via adiabatic evolution. The small gap arises from locally independent choices, which cause the system to initially evolve and localize into a configuration far from the solution in terms of Hamming distance. We investigate remedies to this problem. Specifically, we show that quantum quenches in these models can exhibit signatures of quantum many-body scars, which in turn, can circumvent the superexponential gaps. By quenching from a suboptimal configuration, states with a larger ground state overlap can be prepared, illustrating the utility of quantum quenches as an algorithmic tool.
\end{abstract}

\maketitle 

\section{Introduction}

\begin{figure}[H]
    \includegraphics[width=.99\linewidth]{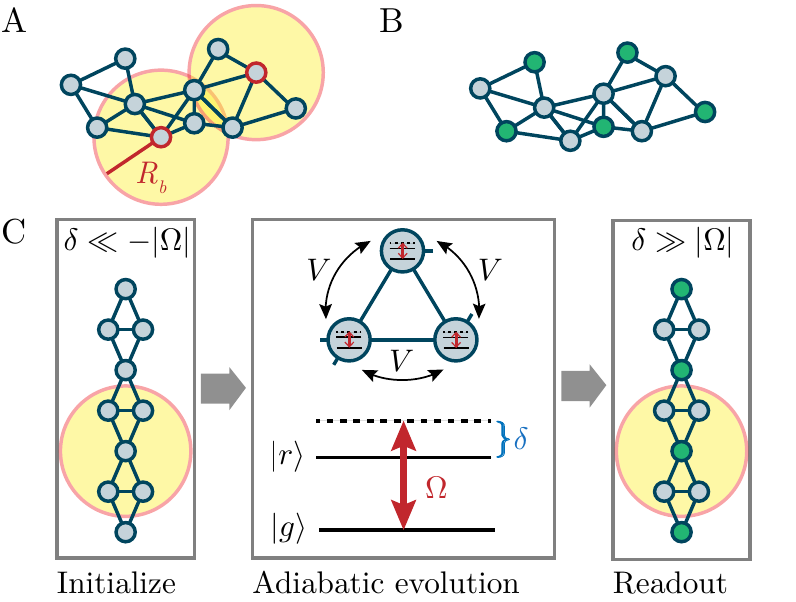}
    \caption{\textbf{(A)}~A unit-disk graph, where all pairs of vertices separated by less than $R_b$ share an edge. \textbf{(B)}~A solution of the maximum independent set (MIS) problem with green vertices constituting the MIS. \textbf{(C)}~Schematic of steps to solve the MIS problem on a Rydberg atom array. The graph is initialized with all atoms in the atomic ground state $\ket{gg\dots g}$ with connected sites located inside their respective Rydberg blockade radii. The parameters of the adiabatic evolution are the detuning $\delta$, the Rabi frequency $\Omega$, and the interaction  $V$. After a sufficiently slow sweep from $\delta \ll -|\Omega|$ to $\delta \gg |\Omega|$, the set of sites in the state $\ket{r}$ corresponds to the MIS.}
    \label{fig:Fig1}
\end{figure}

Adiabatic quantum evolution offers a resource-efficient way to explore the power of current quantum computing platforms. While adiabatic quantum computation is equivalent to circuit-based computation in terms of computational complexity~\cite{Aharonov2007Adiabatic}, it incurs smaller experimental overheads provided classical problems can be encoded in quantum Hamiltonians in a hardware-efficient manner. 
Recently, Rydberg atom arrays have emerged as an especially promising platform for both analog and digital quantum information processing~\cite{Henriet2020Quantum, Ebadi2020Quantum, Scholl2021Quantum, Bluvstein2021Controlling, Ebadi2022Quantum}. The Hamiltonian of this system naturally encodes the so-called maximum independent set (MIS) problem (Fig.~\ref{fig:Fig1}), which is NP-hard in the worst case~\cite{Clark1990Unit}, and has many real-word applications ranging from map labeling~\cite{Verweij1999Optimisation} to  automated DNA design~\cite{Hossain2020Automated}. Using optical tweezers, atoms can be arranged in nearly arbitrary two-dimensional geometries to encode the MIS problem on unit-disk graphs. The adiabatic approach to solving the MIS problem consists of initializing all atoms in the ground state and applying a time-dependent drive to the Rydberg state~\cite{Pichler2018Quantuma, Ebadi2022Quantum}. In the ideal limit, as long as the maximum rate of change of the drive is sufficiently small, the MIS state will be prepared. However, in practice, the required 
evolution time is determined by the smallest energy gap for any given 
instance of the problem.  While recent experimental observations indicate  a superlinear quantum speedup in solving the Maximum Independent Set problem on certain unit-disk graph instances \cite{Ebadi2022Quantum}, it remains an important open challenge to determine sets of instances that are either easy or hard to solve adiabatically and how these compare to instances that are difficult classically. 
Insights into what contributes to the difficulty of instances could enable the development of new algorithms overcoming such algorithmic slowdowns.

\clearpage
In this work, we construct instances of the MIS problem that are very hard to solve via an adiabatic approach. 
Our construction is based on locally independent choices when forming suboptimal independent sets of a certain size. The ability to locally pick different vertices without affecting the rest of the independent set gives rise to a large degeneracy. The coupling between the atomic ground state and superpositions of these degenerate configurations is enhanced, causing the system to initially evolve toward such states during the protocol. The transition into the MIS at later times encounters a bottleneck if the locally degenerate configurations are separated from the MIS by a large Hamming distance. We provide an explicit example of a quasi-one-dimensional chain, where the Hamming distance between the MIS and the degenerate configurations is proportional to the system size. Using a mean-field and perturbation-theory analysis supported by numerics, we show that the ground state undergoes an avoided level crossing whose gap vanishes superexponentially with the system size. The avoided level crossing may be viewed as a first-order transition from the locally degenerate configurations to the MIS.

Related obstructions to the adiabatic algorithm have been previously reported for the weighted MIS, the 3-satisfiability, and the exact cover problems~\cite{Amin2009First,Farhi2010Quantum, Altshuler_2010}. In particular, it was shown~\cite{Altshuler_2010} that random instances of the exact cover problem are likely to give rise to a superexponentially small gap due to Anderson localization in configuration space. Our work demonstrates that such small gaps can occur even in the absence of spatial disorder. Moreover, as locally independent choices create a large degeneracy for independent sets smaller than the MIS, they also affect classical algorithms: the time required to find the MIS using simulated annealing with local updates increases at least exponentially with the system size~\cite{Ebadi2022Quantum}.

We discuss two approaches to address the challenges associated with the small gap. First, we add spin-exchange terms to the Hamiltonian \cite{Cain2022Quantum}, which can be experimentally realized, e.g., using multiple Rydberg levels~\cite{Glaetzle2015Designing,Scholl2022Microwave,Steinert2022Spatially}. These terms allow the system to explore the configuration space more efficiently. We observe that the minimum gap increases by orders of magnitude, although the superexponential scaling persists. As a second remedy, we consider nonadiadbatic quenches, which give rise to quantum many-body scars~\cite{Bernien2017Probing, Serbyn2021Quantum}. These scars lead to oscillations between the locally degenerate configurations and the MIS, and thus offer an avenue to preparing the MIS without having to adiabatically cross the first-order transition. Indeed, we find significantly improved probablilities for obtaining the MIS solution.  Further, we show that the oscillations decay with a slow  exponent,  and hence,  provide a scaling advantage over a superexponential adiabatic slowdown.

First, let us give a brief overview of the physical models that we consider in our analysis. Our starting point is a system of Rydberg atoms arranged in a chain of diamonds as in Fig.~\ref{fig:Fig1}(C), which we call the \emph{doublet Rydberg model}. In this model, the atoms are coupled by the long-ranged van der Waals interactions. In the limit of strong interactions, the simultaneous excitation of neighboring atoms to the Rydberg state is strongly suppressed, leading to the so-called PXP Hamiltonian, where adjacent excitations are strictly forbidden and the tails of the van der Waals interactions are ignored~\cite{Lesanovsky2012Interacting}. Given the particular geometry of the doublet chain, we call this the \emph{doublet PXP model}. We show that for the adiabatic algorithm, the doublet PXP model is equivalent to a one-dimensional PXP chain wherein every odd site has its driving frequency enhanced by a factor of $k$. We refer to this imbalanced PXP model as the \emph{$k$-PXP chain}. A related model, the \emph{$k$-Rydberg chain}, is obtained by adding back the full van der Waals interaction to a chain of atoms where the driving term on every odd site is enhanced by a factor $k$.

This paper is organized as follows. In Sec.~\ref{sec:MIS}, we review the MIS problem and define the quantum adiabatic algorithm in the context of Rydberg atom arrays. Next, in Sec.~\ref{sec:Q1D}, we describe the construction of graphs with locally independent choices and introduce the quasi-one-dimensional chain forming a lattice of interconnected diamond-shaped plaquettes. Thereafter, in Sec.~\ref{sec:Localization}, we analyze the properties of this chain of diamonds in detail and show that the minimum gap vanishes superexponentially with the system size. We investigate remedies to overcome the unfavorable scaling of the gap---either by adding hopping terms to the Hamiltonian in Sec.~\ref{ssec:flipflop} or by exploiting quantum many-body scars following a quench in Secs.~\ref{ssec:scars}~and~\ref{ssec:2D}---before concluding in Sec.~\ref{sec:Conclusion}.

\section{The Maximum Independent Set problem in Rydberg atom arrays} \label{sec:MIS}

We start by reviewing the problem Hamiltonian and the adiabatic algorithm. 
The combinatorial problem of interest here is the MIS problem. 
Given a graph $G(\mathcal{V}, \mathcal{E})$ with vertices $\mathcal{V}$ and edges $\mathcal{E}$, an independent set is a subset of vertices in which no two vertices share an edge. A \emph{maximum} independent set is an independent set with the largest possible number of vertices. There may be several distinct MIS configurations; the MIS problem is to find at least one such MIS. The MIS problem may equivalently be expressed as the problem of finding the ground state of the classical spin Hamiltonian 
\begin{align}
    H^{}_\text{MIS} = \sum_{(i,j) \in \mathcal{E}} V n^{}_i n^{}_j - \delta \sum_{i \in \mathcal{V}} n^{}_i,
\end{align}
for $V$\,$>$\,$\delta$\,$>$\,$0$. Here, we associate with each site $i$\,$\in$\,$\mathcal{V}$ a two-level system with computational basis states $\{\ket{0}, \ket{1}\}$. The Hamiltonian is expressed in the occupation-number basis $n_i = \ket{1}_i\bra{1}$, and we will refer to $V$ and $\delta$, respectively, as the interaction strength and the detuning. An MIS is identified with a ground state of this Hamiltonian by preparing all vertices in the independent set in the $\ket{1}$ state and all others in $\ket{0}$.

We restrict the graphs of interest to unit-disk graphs, which are constructed by placing vertices in a 2D plane and connecting each vertex to any other within a fixed radius; see Fig.~\ref{fig:Fig1}(A) for an example. Note that the MIS problem remains NP-complete when restricted to unit-disk graphs~\cite{Clark1990Unit}. It was observed in Ref.~\onlinecite{Pichler2018Quantuma} that for the special class of unit-disk graphs, an approximation to $H_\text{MIS}$ is natively realized in Rydberg platforms, which allows one to solve the MIS problem by preparing the ground state of an array of Rydberg atoms. We associate with $\ket{0}$ the ground state of an atom and with $\ket{1}$ a highly excited Rydberg state. For convenience, we will also refer to these states as $\ket{g}$ and $\ket{r}$, respectively. The Rydberg Hamiltonian is given by
\begin{align}
\label{eq:HRyd}
    H^{}_\text{Ryd}(\delta, \Omega) = &\sum_{i, j \in \mathcal{V}} \frac{V}{|r^{}_i-r^{}_j|^6} n^{}_i n^{}_j - \delta \sum_{i \in \mathcal{V}} n^{}_i 
    + \Omega \sum_{i \in \mathcal{V}} \sigma_i^x,
\end{align}
where $\Omega$ denotes the global Rabi frequency. 
Here, the interaction along the edges of the graph has been replaced by a term that depends on the sixth power of the Euclidean distance between two atoms (vertices), arising from the van der Waals (vdW) interaction between pairs of atoms occupying the Rydberg state. 
The first two terms in Eq.~\eqref{eq:HRyd} encode the classical cost function whereas the last one is the quantum driver, which is necessary to adiabatically prepare the MIS.

In the quantum adiabatic algorithm (QAA)~\cite{Messiah1962Quantum, Farhi2000Quantum}, the system is prepared in the initial state $\ket{gg\dots g}$, which is the (approximate) ground state of $H_0 = H_\text{Ryd}(\delta \ll -|\Omega|, \Omega)$. The parameters of the Hamiltonian are then slowly changed until the target Hamiltonian $H_T=H_\text{Ryd}(\delta \gg |\Omega|, \Omega)$ is reached. We will assume that this change proceeds according to the linear interpolation
\begin{align}
    H\left(\frac{t}{T}\right) = \left( 1-\frac{t}{T}\right) H_0 + \frac{t}{T} H_T,
\end{align}
where $t$ is the time and $T$ the total duration~\footnote{In practice, the Rabi frequency $\Omega$ is usually ramped from zero to a constant value in the beginning of the sweep and ramped down at the end. For simplicity of the presentation, we assume a constant $\Omega$ throughout the sweep.}. The preparation time $T$ required to achieve an appreciable overlap with the ground state of $H_T$ is related to the adiabatic gap $\Delta_\text{min}$, which is the minimum spectral gap between the ground state and the first excited state during the evolution. In the worst case, the adiabatic theorem gives a scaling of $T$  with the adiabatic gap as $1/\Delta_\text{min}^3$~\cite{Jansen2007Bounds} such that a small gap may render the adiabatic algorithm highly inefficient in practice.

\section{Locally independent choices in quasi-1D graphs} \label{sec:Q1D}
Recently,~\citet{Ebadi2022Quantum} observed that for a class of random unit-disk graphs, degeneracies play an important role in determining the instance-to-instance variation in the hardness of solving the combinatorial optimization problem. Specifically, we define the degeneracy ratio 
\begin{align}
    R \equiv D(|\text{MIS}|-1) / D(|\text{MIS}|),
\end{align}
where $D(s)$ denotes the number of distinct independent sets of size $s$ and $|\text{MIS}|$ is the size of the MIS. Reference~\onlinecite{Ebadi2022Quantum} reported that $R$ is strongly correlated with the hardness of many graph instances for both simulated annealing and QAA~\footnote{In Ref.~\onlinecite{Ebadi2022Quantum}, a hardness parameter is defined as $\mathcal{HP}=R/|\text{MIS}|$. However, $|\text{MIS}|$ scales at most linearly with the system size, whereas $R$ scales exponentially for the chain of diamonds. The denominator in the hardness parameter is therefore a subleading contribution and can be safely ignored for the purpose of this work.}. In the case of simulated annealing, this is because it is challenging to find an independent set that is connected to the MIS via local updates if the degeneracy ratio is large such that the algorithm will typically get stuck in a local minimum. For  QAA, the relation between a large degeneracy ratio and a long computation time---or equivalently, a small minimum gap---is less clear. In this section, we present a mechanism that simultaneously gives rise to a large degeneracy ratio and a small adiabatic gap.

\begin{figure}[t]
    \includegraphics[width=1.0\linewidth]{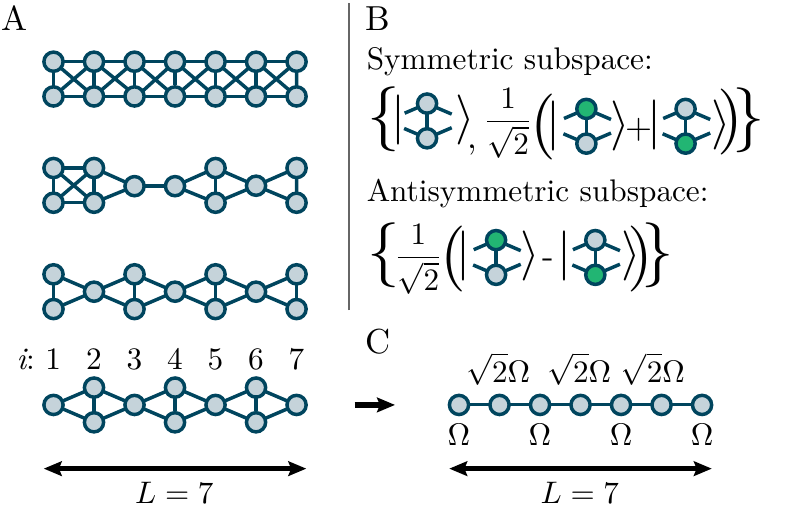}
    \caption{\textbf{(A)}~Different quasi-1D chains. In this work, we focus mainly on the fourth graph (chain of diamonds). Singlets and doublets are indexed in the horizontal direction. \textbf{(B)}~In the limit of strong blockade, two sites sharing a vertical edge (doublet) may be decomposed into the symmetric and antisymmetric subspace in the occupation basis. An adiabatic sweep from the $\ket{gg\dots g}$ state is restricted to the symmetric subspace for every doublet. \textbf{(C)}~As the antisymmetric subspace is effectively frozen out, the quasi-1D chains can be mapped to a regular 1D chain with site-dependent Rabi frequencies, shown here  for the chain of diamonds. The doublet $(i,j)$, where each site contributes to the Hamiltonian with a local drive $\Omega ( \sigma_i^x +  \sigma_j^x)$, is replaced by a single node with an enhanced Rabi frequency $\sqrt{2}\Omega$.}
    \label{fig:Fig2}
\end{figure}

The mechanism is based on local degrees of freedom that give rise to a large degeneracy. We start by considering quasi-1D graphs, although this mechanism can be generalized to higher dimensions. The graphs of interest are built from a finite 1D  chain by replacing some of the vertices with two connected vertices (a doublet). Four different examples with a chain of length $L$\,$=$\,$7$ are shown in Fig.~\ref{fig:Fig2}(A). The fourth graph (chain of diamonds) and the third graph (diamonds with doublets on the boundaries) are special in the sense that they have a particularly large or small degeneracy ratio when $L$ is odd. For the chain of diamonds, there is only a single MIS solution [Fig.~\ref{fig:Fig1}(C)]. The number of MIS-1 configurations, i.e., independent sets that are smaller than an MIS by one vertex, is exponential in the chain length $L$ because of the freedom to occupy either position in every doublet. Hence, the degeneracy ratio $R$ increases exponentially with $L$. In the remainder of this paper, we focus on the fourth graph in Fig.~\ref{fig:Fig1}(A), referring to it as the \emph{doublet chain}.

For encoding these graphs in Rydberg atom arrays, we shall consider the regime of Rydberg blockade~\cite{jaksch2000fast}, where the interaction strength between the two atoms in each doublet is sufficiently large so that they cannot simultaneously occupy the Rydberg state. Here, we do not consider longer-range interaction tails yet. In this case, the states of a doublet are spanned by the symmetric subspace $\{\ket{gg},(\ket{gr} +\ket{rg})/\sqrt{2}\}$ and the antisymmetric subspace $\{(\ket{gr} - \ket{rg})/\sqrt{2}\}$ [see Fig.~\ref{fig:Fig2}(B)]. As the linear, homogeneous adiabatic sweep is initiated in the $\ket{gg\dots g}$ state, the antisymmetric subspace is never populated. The symmetric subspace may be viewed as a two-level system, for which the local Rabi frequency is enhanced by a factor of $\sqrt{2}$ compared to a single atom. This allows us to map the \emph{doublet PXP model} to a 1D chain with alternating Rabi frequencies [Fig.~\ref{fig:Fig2}(C)].
For this 1D geometry, we consider both the strongly interacting limit (\emph{$\sqrt{2}$-PXP chain}) and a finite interaction strength $V$ between even and odd sites (\emph{$\sqrt{2}$-Rydberg chain}). Longer-ranged interactions are referred to as interaction tails and studied only in the \emph{$\sqrt{2}$-Rydberg model} when explicitly indicated.

The enhancement of the effective Rabi frequency for doublets excites them to the Rydberg state faster than atoms that do not belong to a doublet. In the picture where the doublet chain is viewed as a 1D chain with alternating Rabi frequencies, the system is driven toward a state where the even sites are excited to a Rydberg state, while the odd sites remain in the ground state. Since $L$ is odd, this state belongs to the MIS-1 subspace. We will show below that to transition from this state to the MIS, where the odd sites occupy the Rydberg state, it is necessary to pass through an avoided crossing whose gap decays superexponentially with $L$. Hence, the locally independent choices under the independent set constraint result in a small adiabatic gap by causing the system to initially favor a suboptimal solution.

\section{Local degrees of freedom make the adiabatic algorithm fail} \label{sec:Localization}
We now turn to investigate the particular properties of the adiabatic gap for the $\sqrt{2}$-models, where every even site has an enhanced Rabi frequency.

\subsection{Phase diagram for varying imbalance strength} \label{ssec:PD}
\begin{figure}[tbp]
    \includegraphics[width=1.0\linewidth]{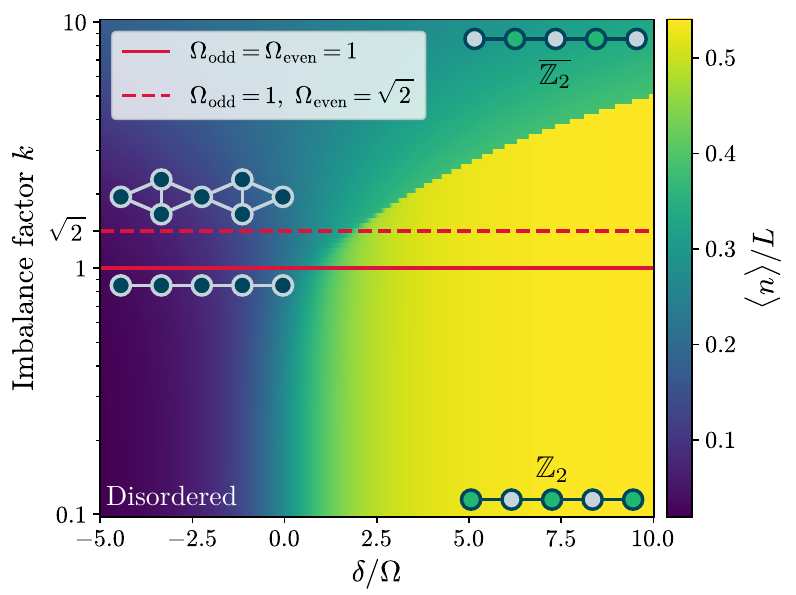}
    \caption{The mean occupation for a $k$-PXP chain of length $L=11$ with the Rabi frequency enhanced by a factor $k$ on every even site. The doublet chain corresponds to $k=\sqrt{2}$. The adiabatic sweeps considered originate on the left side for $\delta /\Omega \ll -1$ and end on the right at $\delta /\Omega \gg 1$. The transition from the disordered to the ordered phase along horizontal lines becomes sharper for larger $k$. Notably, the MIS solution (denoted by $\mathbb{Z}_2$) becomes separated by a first-order transition from the phase corresponding to the MIS-1 solution ($\overline{\mathbb{Z}_2}$) when $k>1$. Different $k$ may be realized by other gadgets replacing the doublets (see Appendix~\ref{app:gadget}).}
    \label{fig:Fig3}
\end{figure}

In the construction of the doublet chain, we have so far only introduced $\Omega_\text{even} = \sqrt{2} \Omega_\text{odd}$ (i.e., $k$\,$=$\,$\sqrt{2}$). However, instead of the doublet, other gadgets may be used~\cite{Nguyen2022Quantum} that replace a single site in the 1D chain; e.g., a fully connected graph of 3 vertices that all connect to the previous and subsequent vertex. This gadget would lead to a local enhancement of $k$\,$=$\,$\sqrt{3}$ (as discussed in App.~\ref{app:gadget}). To build intuition for the imbalanced chain, we construct a phase diagram of the system as a function of $k$ and the detuning $\delta$, using the mean occupation as an observable~\cite{PhysRevLett.124.103601,Kalinowski2022Bulk}. We find a disordered ground state for $\delta/\Omega \ll -1$ and two ordered ground states for $\delta/\Omega \gg 1$ in Fig.~\ref{fig:Fig3}. The red horizontal solid and dashed lines correspond to adiabatic sweeps for two particular geometries, i.e., the PXP chain and the $\sqrt{2}$-PXP chain, respectively. We denote one ordered state as $\mathbb{Z}_2$ (yellow) corresponding to the MIS solution. In the other ordered state, the sites occupied are exactly those which are unoccupied in $\mathbb{Z}_2$; thus, we label it by $\overline{\mathbb{Z}_2}$ (green). From the figure, we observe that the transition into the $\mathbb{Z}_2$ phase becomes increasingly sharper for larger $k$. For $k$\,$=$\,$1$, the model is known to exhibit a translational-symmetry-breaking second-order phase transition equivalent to that of the transverse-field Ising model~\cite{Fendley2004Competing}. Away from the $k$\,$=$\,$1$ line, the $\mathbb{Z}_2$ translational symmetry is explicitly broken in the Hamiltonian by the staggered Rabi frequencies, so there is no sharp distinction between the $\mathbb{Z}_2$, $\overline{\mathbb{Z}_2}$, and disordered states in terms of their symmetries. We observe that the enhanced Rabi frequency on alternate sites now leads to the occurrence of a first-order transition for $k$\,$\gg$\,$1$, while the two states can be continuously connected in the $k$\,$\ll$\,$1$ regime. This line of first-order transitions that terminates at a second-order critical point is reminiscent of the familiar liquid--gas transition in water, which is similarly characterized by a first-order transition without any symmetry breaking~\cite{jimenez2021quantum}.

\begin{figure}[tbp]
    \includegraphics[width=1.0\linewidth]{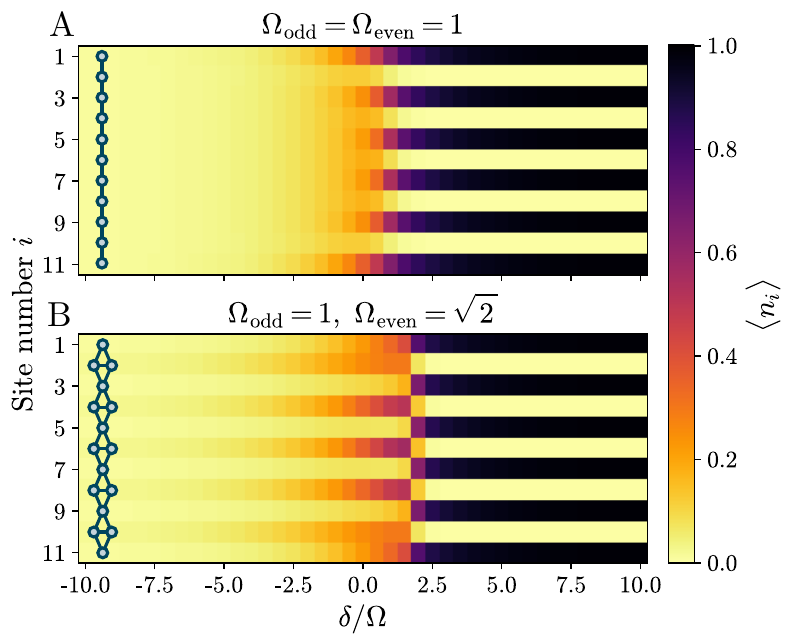}
    \caption{Local occupation $\langle n_i\rangle$ in the ground state at different values of $\delta /\Omega$ for \textbf{(A)} a Rydberg chain and \textbf{(B)} a $\sqrt{2}$-Rydberg chain ($V=100$, no interaction tails) of length $L=11$. For $\delta /\Omega \ll (\delta/\Omega)_\text{crit}$ the local occupation is zero for both chains and converts into a stripe pattern ($\mathbb{Z}_2$) for $\delta /\Omega \gg (\delta/\Omega)_\text{crit}$. However, close to the respective critical points, the behavior is different. The Rydberg chain shows an onset of increasing occupation at the boundary first and in the bulk only slightly later. The $\sqrt{2}$-Rydberg chain shows increased occupation on the doublet sites before the critical point. In an adiabatic sweep, the system will initially rotate into the wrong configuration ($\overline{\mathbb{Z}_2}$), leading to a very small adiabatic gap.}
    \label{fig:Fig5}
\end{figure}

Let us outline a general explanation of the first-order transition in $\sqrt{2}$-chains before presenting analytical arguments using both a mean-field picture and a perturbation-theory calculation in the following subsections.
Intuitively, the enhanced Rabi frequency locally favors occupation on the even sites. This is visualized clearly in Fig.~\ref{fig:Fig5} when comparing the local Rydberg occupation $\langle n_i \rangle$ of the Rydberg chain in (A) to that of the $\sqrt{2}$-Rydberg chain in (B) for varying $\delta/\Omega$. Approaching the critical detuning from negative to positive values of $\delta/\Omega$, we observe increased occupation on the even sites of the quasi-1D chain. From the perspective of the MIS problem, the \emph{wrong} sites are favored and rotate first, which is not the case for the second-order phase transition in the 1D chain. Hence, for the$\sqrt{2}$-chain, the ground states shortly before the critical point ($\overline{\mathbb{Z}_2}$) and shortly after ($\mathbb{Z}_2$) have near-maximal Hamming distance. This leads to a very small adiabatic gap because high-order quantum fluctuations are required to transition between the two states. We emphasize an interplay between the bulk and the boundary in the ordering process: in Fig.~\ref{fig:Fig5}(A), we observe that the boundary orders first, thereby assisting the adiabatic evolution~\cite{Kalinowski2022Bulk}. This is primarily due to the presence of fewer interactions at the boundary, which favors vertices close to the ends of the chain to be occupied earlier. While this phenomenon can also be seen for the quasi-1D chain, it is much weaker than the effect of the local enhancement of the Rabi frequency. 

\subsection{Mean-field analysis}
\label{sec:MF}
We now establish a mean-field description of the $\sqrt{2}$-PXP chain to phenomenologically describe the features of the graph geometry. To this end, we consider a simple product-state ansatz, which obeys the blockade condition for the limit of large interactions. For the ground state and the first excited state, we write
\begin{align}
    \ket{\psi^{}_{\mathbb{Z}_2}} &= \ket{\phi_1} \ket{g} \ket{\phi_3} \ket{g} \cdots \ket{\phi_L},\\
    \ket{\psi^{}_{\overline{\mathbb{Z}_2}}} &= \ket{g} \ket{\phi_2} \ket{g} \ket{\phi_4} \cdots \ket{g},
\end{align}
with $\ket{\phi_i} \equiv \alpha_i \ket{g} + \beta_i \ket{r}$. Next, we assume that all odd and even sites can be described by the same two-level dynamics such that $\ket{\phi_1} = \ket{\phi_3} = \dots = \ket{\phi_\text{odd}}$ and $\ket{\phi_2} = \ket{\phi_3} = \dots = \ket{\phi_\text{even}}$.\\
The local Hamiltonian is given by 
\begin{align}
    H_i = -\delta\; n_i + 
    \begin{cases}
    \Omega\; \sigma_i^x, & i\ \text{odd},\\
    k \Omega\; \sigma_i^x, & i\ \text{even},
    \end{cases}
\end{align}
and the local state follows as $\ket{\phi_i}$\,$=$\,$ \cos(\theta_i/2) \ket{r} + \sin(\theta_i/2) \ket{g}$ with $\tan(\theta_\text{odd}) = -2\Omega/\delta$ and $\tan(\theta_\text{even}) = -2k\Omega/\delta$. With this ansatz, the energies of the ground and first excited states are 
\begin{align}
    E^{}_{\mathbb{Z}_2}&=-\frac{L+1}{4}\left(\delta+\sqrt{4\Omega^2+\delta^2}\right),\quad \text{and}\\
    E^{}_{\overline{\mathbb{Z}_2}} &= -\frac{L-1}{4}\left(\delta+\sqrt{4(k\Omega)^2+\delta^2}\right),
\end{align}
respectively. Indeed, the mean-field picture explains Fig.~\ref{fig:Fig5}(B), where on the two sides just next to the critical point, the two states are rotated in a way so as to yield complementary stripe patterns. Furthermore, we may look for the critical point where the two energies $E_{\mathbb{Z}_2}$ and $E_{\overline{\mathbb{Z}_2}}$ would cross in the mean-field model. Setting $k=\sqrt{2}$, we find that $(\delta/\Omega)_\text{crit} \sim \mathcal{O}(\sqrt{L})$, implying that when the $\sqrt{2}$-PXP chain growths in length, the avoided level crossing occurs in an increasingly classical regime.

\subsection{Superexponentially small adiabatic gap}

Having established an intuitive understanding of the doublet chain from the phase diagram and the mean-field model, we now analyze the scaling of the minimal spectral gap in an adiabatic evolution process in which the detuning $\delta$ is swept from large negative to large positive values. To anticipate the result, we find that the adiabatic gap vanishes superexponentially fast with growing system size.

\begin{figure}[tb]
    \includegraphics[width=1.0\linewidth]{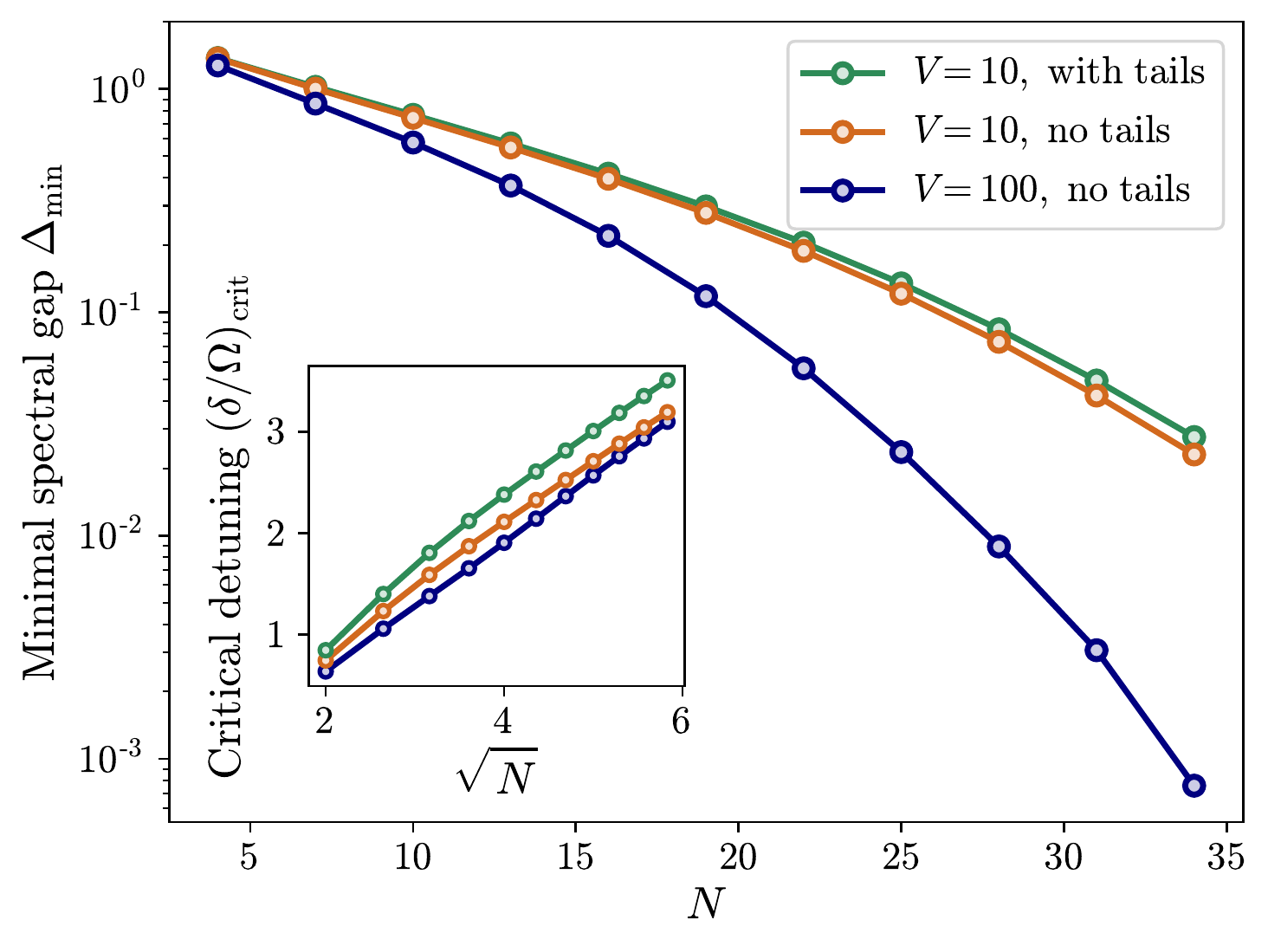}
    \caption{Size and position of the adiabatic gap in the $\sqrt{2}$-Rydberg chain. The main plot shows the superexponential scaling of the adiabatic gap for different interaction strengths. Adding the tails of the Rydberg interaction decaying as $1/r^6$ to the $\sqrt{2}$-Rydberg chain does not change the slope significantly. In the inset, we show that the value of the critical detuning $(\delta/\Omega)_\text{crit}$ grows with the square root of the system size, confirming the mean-field prediction. Hence, the phase transition occurs in an increasingly classical regime for long chains.}
    \label{fig:sizeposition}
\end{figure}

First, we discuss the numerical results for $k$\,$=$\,$\sqrt{2}$ and then follow up with an analytical proof for sufficiently large $k$. The Hamiltonian for the doublet chain reduced to a $\sqrt{2}$-Rydberg chain is given by
\begin{align}
    H^{}_\text{Q1D} &= \sum_{i>j} \frac{V}{|i-j|^6} n^{}_i n^{}_j - \delta \sum_i^L n^{}_i \\
    & + \Omega \sum_{i \; \text{odd}}^L \sigma_i^x + \sqrt{2} \Omega \sum_{i \; \text{even}}^L\sigma_i^x.
\end{align}
In Fig.~\ref{fig:sizeposition}, we compute the size and position of the adiabatic gap and also compare the effect of varying the strength of the interaction $V$ as well as the inclusion of long-ranged vdW tails in the $\sqrt{2}$-Rydberg chain up to a distance of five sites. The Rabi frequency is fixed to $\Omega=1$, such that the only parameter swept in the adiabatic evolution is the detuning from $\delta$\,$\ll$\,$-1$ to $\delta$\,$\gg$\,$1$. The adiabatic gap is computed using the density-matrix renormalization group (DMRG) algorithm~\cite{Schollwoeck2011density} implemented in ITensor~\cite{Fishman_2022}. Our numerical data clearly shows that the size of the adiabatic gap approaches zero with a scaling that is superexponential, even in the presence of long-ranged interaction tails or a softer blockade between horizontal nodes. Furthermore, we numerically confirm the mean-field prediction that the position of the spectral gap shifts toward larger values of $\delta$ and scales with the square root of the chain length.

Now, we seek to confirm the numerical findings with an analytical description of the gap's scaling. In order to do so, we present a three-step argument summarizing a perturbative approach and include a full description thereof in the Appendix. The argument is as follows:
\begin{itemize}
    \item[1.] From the mean-field model, we compute that the position of the critical point lies in the classical regime with $\delta_\text{crit} \gg |\Omega|$ for large $L$. Hence, a perturbative ansatz is warranted.
    \item[2.] Next, we consider the $k$-PXP chain with locally enhanced Rabi frequencies, described by the Hamiltonian $H/\delta = H_0+\Omega/\delta\; H_x$\,$\equiv$\,$ H_0$\,$+$\,$\epsilon H_x$. Here, the unperturbed Hamiltonian is $H_0$\,$=$\,$- \sum_i^{L} n_i$, and the perturbation is
    \begin{align*}
        H_x = \sum_{i \; \text{odd}}^{L} P^{}_{i-1}\sigma_i^x P^{}_{i+1} + k \sum_{i \; \text{even}}^{L} P^{}_{i-1}\sigma_i^x P^{}_{i+1},
    \end{align*}
    where the projectors $P_i \equiv (1-n_i)$ and $P_0=P_{L+1}=I$ enforce the blockade constraint. Now, considering the subspace of the first excited states of $H_0$ and using Rayleigh-Schrödinger perturbation theory, the lowest perturbed eigenenergy is found for the eigenstate that completely localizes in the $\overline{\mathbb{Z}_2}$ state if the local enhancement factor $k$ is sufficiently large.
    \item[3.] For sufficiently large $k$, the spectral gap at the avoided level crossing is given, similar to a Landau-Zener model~\cite{Landau1932Zur}, as 
    \begin{align*}
        \Delta \propto \epsilon_\text{crit}^L = \sqrt{1/L}^{L} = \exp(-L\log(L)/2).
    \end{align*}
\end{itemize}
This agrees with the numerical findings of a superexponentially small adiabatic gap as the length of the doublet chain is increased.

\section{Mitigating the small adiabatic gap} \label{sec:remedies}

In the following section, we investigate two approaches to overcome the superexponentially small gap encountered in preparing the MIS ground state with quantum dynamics.

\subsection{Additional spin-exchange terms in the evolution} \label{ssec:flipflop}

\begin{figure}[tb]
    \includegraphics[width=1.0\linewidth]{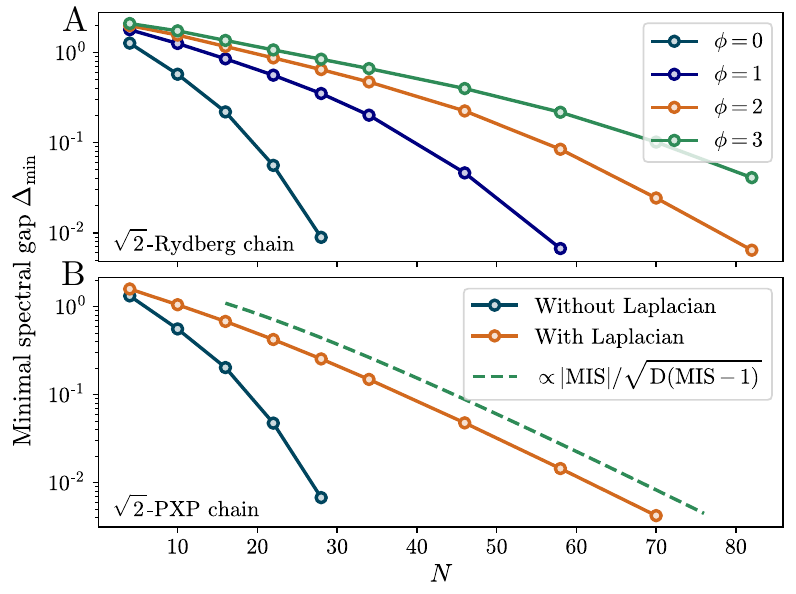}
    \caption{\textbf{(A)} Adiabatic gap with an additional spin-exchange term of varying global strength $\phi$ in the  $\sqrt{2}$-Rydberg chain ($V$\,$=$\,$100$, no tails). The $x$-axis shows the number of sites $N$ in the corresponding quasi-1D geometry. We observe that the superexponential scaling persists. However, the absolute size of the gap increases by orders of magnitude for the longest chains probed. \textbf{(B)}~Comparison between the gap in a standard adiabatic sweep and the Laplacian formalism. Here, we consider the $\sqrt{2}$-PXP model by setting the interaction strength to a very large value ($V$\,$=$\,$1000$). The strength of the Laplacian was set to $\phi=L$. The expected scaling, shown as a guide, becomes exponential in the limit of $N\gg1$.}
    \label{fig:Fig6}
\end{figure}

The first approach is to add additional terms to the Hamiltonian  that facilitate fluctuations between the $\mathbb{Z}_2$ and the $\overline{\mathbb{Z}_2}$ state. Intuitively, the addition of boson hopping, or spin-exchange, terms, $\sigma_i^+\sigma_j^- + \sigma_i^-\sigma_j^+ = (\sigma_i^x\sigma_j^x+\sigma_i^y\sigma_j^y)/2$ with $\sigma_i^\pm = (\sigma_i^x \pm i\sigma_i^y)/2$, allows for occupations to move across the chain directly as opposed to via higher-order processes with the $\sigma_i^x$ term alone. To confirm this hypothesis, we numerically compute the adiabatic gap for $\sqrt{2}$-Rydberg chains of lengths up to $L=55$, corresponding to $N=82$ sites in the Rydberg model. The Hamiltonian, including the spin-exchange terms, that we consider is
\begin{align}\label{eq:SE}
    H^{}_\text{SE} = H^{}_\text{Q1D} - \phi \sum_{i=1}^{L-1} (\sigma^+_i\sigma^-_{i+1} + \sigma^-_i\sigma^+_{i+1}).
\end{align}
We probe values of $\phi\in\{0,1,2,3\}$, do not include long-ranged tails here, and set $V$\,$=$\,$100$, $\Omega$\,$=$\,$1$. As previously, the only parameter that we sweep in the adiabatic evolution is the detuning from $\delta \ll -1$ to $\delta \gg 1$.  We observe that the scaling of the vanishing adiabatic gap remains superexponential $\Delta_\text{min} \sim \mathcal{O}(\exp(-N \log(N))$ even with the spin-exchange terms. This is expected as the critical region still lies in the classical regime and transitioning into the ground state, as before, requires quantum fluctuations of the order of the chain length. Importantly, however, the absolute values of $\Delta_\text{min}$ are significantly larger when the spin-exchange term are present, up to multiple orders of magnitude as visualized in Fig.~\ref{fig:Fig6}(A). We note that Eq.~\eqref{eq:SE}, which describes the $\sqrt{2}$-Rydberg chain, is not equivalent to adding spin-exchange terms of global strength $\phi$ to the doublet Rydberg model.

In a similar vein, Ref.~\onlinecite{Cain2022Quantum} proposes a modification to the original Hamiltonian by adding a global Laplacian term that favors delocalization in configuration space. This graph Laplacian resembles $H_\text{SE}$, but also includes a diagonal correction term that counts the number of possible spin-exchanges. In the $\sqrt{2}$-Rydberg chain, the full Hamiltonian with this modification reads
\begin{align}\label{eq:Laplacian}
    H^{}_\text{L} &= H^{}_\text{Q1D} - \sqrt{2}\phi \sum_i^{L-1} (\sigma^+_i\sigma^-_{i+1} + \sigma^-_i\sigma^+_{i+1}) \\
& + \sum_{i=1}^{L-1} \phi^{}_i n^{}_iP^{}_{i+1}P^{}_{i+2} + \sum_{i=2}^{L} \phi^{}_i P^{}_{i-2}P^{}_{i-1}n^{}_i,
\end{align}
where $\phi_i$\,$=$\,$\phi$ for odd $i$, $\phi_i$\,$=$\,$2\phi$ for even $i$, and $P_i$\,$=$\,$I$ for $i\notin [1,L]$. For large systems, the adiabatic gap in the presence of a Laplacian of sufficient strength is predicted to be $\Delta_\text{min} $\,$\propto$\,$|\text{MIS}| / \sqrt{D(|\text{MIS}|-1)}$ as the degeneracy of the MIS is one. For a quasi-1D doublet chain of odd length $L$, there is only a single MIS and the number of MIS-1 configurations in the graph is $D(|\text{MIS}|-1)= 2^{(L+3)/2}-(L+5)/2$. Therefore, with the Laplacian modification, the scaling of the adiabatic gap becomes exponential instead of superexponential. We accordingly include a guide to the eye in Fig.~\ref{fig:Fig6}(B), which considers the doublet PXP model in the limit of a strong blockade. With an added Laplacian of strength $\phi$\,$=$\,$L$, the scaling of the numerically computed gap follows the exponential trend in the regime of large system sizes.

\subsection{Using many-body scars to quench across the phase transition} \label{ssec:scars}

In the previous section, we described how the small adiabatic gap of the doublet Rydberg model can be mitigated by additional terms in the Hamiltonian. Ideally though, we would like to find an algorithmic procedure that circumvents the superexponential scaling of the adiabatic gap by using only the Rydberg Hamiltonian. Here, we describe how nonergodic eigenstates in this model can be used to cross the observed first-order transition and provide an algorithmic advantage.

Quantum many-body scars, which have attracted much interest in recent years, describe a phenomenon in which a weak breaking of ergodicity gives rise to coherent state revivals of an initial state after a global quench.
This behavior was first observed in experiments on 1D Rydberg chains, which showed persistent oscillations between the $\overline{\mathbb{Z}_2}$ state and the $\mathbb{Z}_2$ state after a global quench~\cite{Bernien2017Probing}.
Since then, numerous theoretical explanations have emerged for such scars, including nonthermal eigenstates in the spectrum~\cite{Turner2018Weak}, and proximity to models with unstable periodic orbits~\cite{Ho2019Periodic}.
For a theoretical introduction on many-body scars, we direct the interested reader to Ref.~\onlinecite{Serbyn2021Quantum}.

\begin{figure}[t]
    \includegraphics[width=1.0\linewidth]{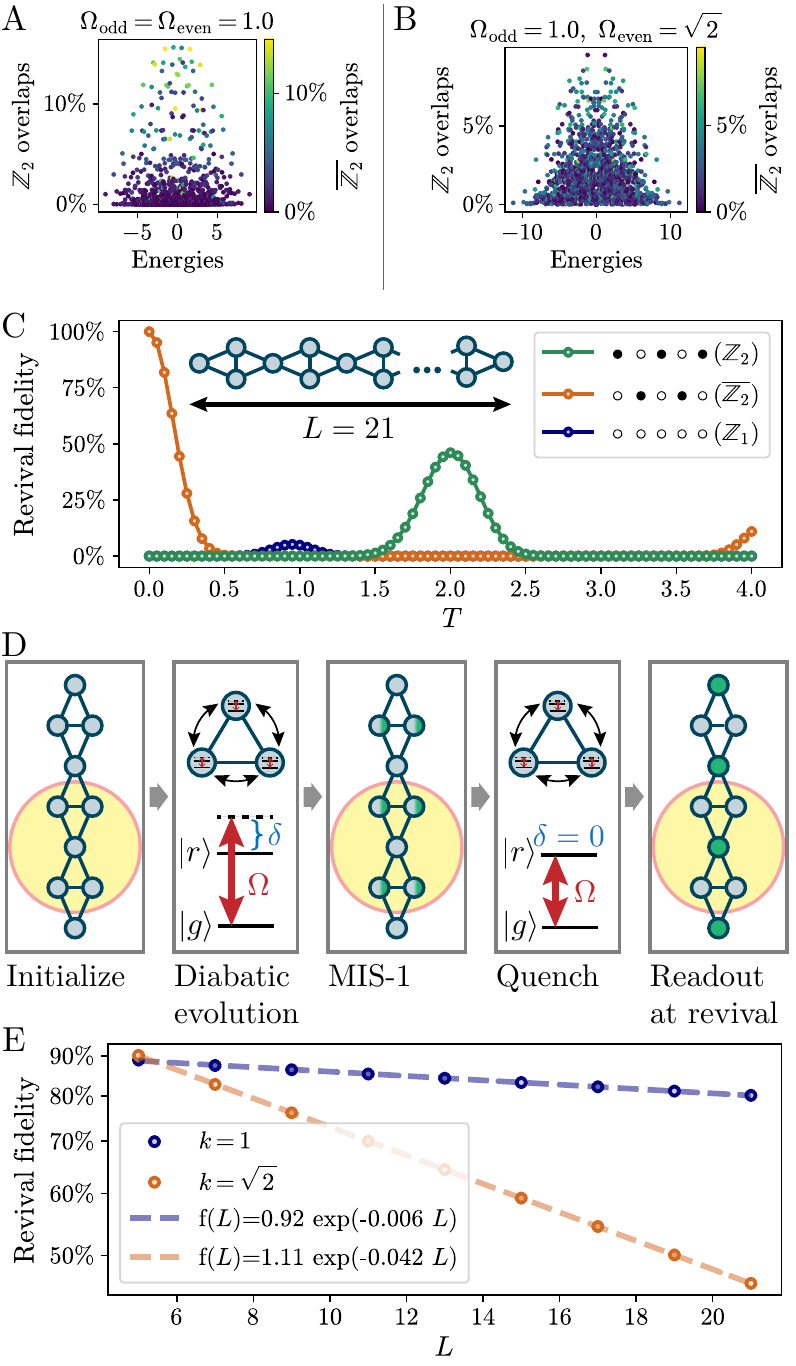}
    \caption{Eigenvector overlaps for the $k$-PXP chain with $\delta=0$ and $L$\,$=$\,$15$ with \textbf{(A)} $k=1$ and \textbf{(B)} $k=\sqrt{2}$ as a function of the energy. 
    \textbf{(C)} Revival and state-transfer fidelity for a quench from the $\overline{\mathbb{Z}_2}$ state, which is the first excited state of the Hamiltonian at large detunings. At $T$\,$=$\,$2.0$, a large fidelity to prepare the $\mathbb{Z}_2$ (MIS) state is observed.
    \textbf{(D)} Quantum many-body scars as an algorithmic tool: instead of slow adiabatic evolution, the system is driven to the $\overline{\mathbb{Z}_2}$ state. Then, after quenching to $\delta=0$, the readout is timed such that the probability to obtain the desired $\mathbb{Z}_2$ state is maximized. 
    \textbf{(E)} Exponential scaling of the peak state-transfer fidelity of the MIS state for the $k$-PXP chains of length up to $L=21$ after a quench to $\delta=0$. The fidelity is extracted at fixed $T$: $T_\text{1D}=2.37$ for the PXP chain and $T_\text{Q1D}=2.00$ for the $\sqrt{2}$-PXP chain. 
    }
    \label{fig:Fig7}
\end{figure}

Since the 1D Rydberg chain is approximated by the a PXP chain, it is natural to ask whether quantum many-body scars also exist in the $\sqrt{2}$-PXP chain and, importantly, whether they provide an alternative route to bypass the first-order transition with global quantum dynamics.
The key signatures of scars are nonthermal eigenstates, which weakly violate the eigenstate thermalization hypothesis (ETH)~\footnote{The violation is weak because the eigenstates constitute a set of measure zero in the thermodynamic limit}. In Fig.~\ref{fig:Fig7}(A,B), we show the overlap of the energy eigenstates of a PXP chain and of a $\sqrt{2}$-PXP chain with the two states $\mathbb{Z}_2$ and $\overline{\mathbb{Z}_2}$ as a function of the energy. 
For the PXP chain [panel (A)], there are signatures of special eigenstates which have much higher overlap with the $\mathbb{Z}_2$ and $\overline{\mathbb{Z}_2}$ states than expected from ETH, as indicated by a gap to the thermal bulk of eigenstates. In contrast, for the $\sqrt{2}$-PXP chain in panel~(B), these features smear out and the special eigenstates start to mix with the bulk of the thermal states.
However, a few eigenstates remain separated, suggesting that the quasi-1D chain may also feature quantum many-body scars, albeit of lower intensity.

To probe quantum scar dynamics, we numerically study the state-transfer fidelity for short times after a global quench to $\delta$\,$=$\,$0$ from the $\overline{\mathbb{Z}_2}$ state [green line in Fig.~\ref{fig:Fig7}(C)] for a chain of length $L$\,$=$\,$21$. The simulations are performed in an effective Hamiltonian subspace where we assume the interaction between vertices connected by an edge to be infinitely large. In the fully blockaded limit, the model is described by a $\sqrt{2}$-PXP chain:
\begin{align}
    H^{}_q = \Omega\sum_{i\ \text{odd}}^L P^{}_{i-1}\sigma_x^iP^{}_{i+1} +  \sqrt{2}\Omega\sum_{i\ \text{even}}^L P^{}_{i-1}\sigma_x^iP^{}_{i+1},
\end{align}
with $P_0$\,$=$\,$P_{L+1}$\,$=$\,$I$. Akin to the PXP chain without enhanced Rabi frequencies, we find significant initial revival fidelities that vanish for later times. Specifically for $L$\,$=$\,$21$, the state-transfer fidelity into the $\mathbb{Z}_2$ state has a peak of 46\% (see Appendix for longer times). From the PXP chain, it is known that the position of the $\mathbb{Z}_2$ peak is approximately independent of system size and is numerically predicted to be at $T_\text{1D}=1.51  \pi/ (2 \Omega) \approx 2.37$~\cite{Bernien2017Probing}. To first order, the revival (and state transfer) periodicity may be approximated by a two-site dimer model which does not take the blockade between two adjacent dimers into account. Such a model predicts $\tilde{T}_\text{1D} = \sqrt{2} \pi/ (2 \Omega) \approx 2.22$, around 6.5\% below the true value. Extending the dimer model to the $\sqrt{2}$-PXP chain of $H_q$, we obtain an approximate prediction $\tilde{T}_\text{Q1D}$\,$=$\,$ \pi/ (\sqrt{3} \Omega)$\,$ \approx$\,$ 1.81$, which is 9.5\% below the numerical prediction of $T_\text{Q1D}=1.27 \pi/ (2 \Omega) \approx 2.00$. The calculation of the revival time is detailed in App.~\ref{app:dimer}.

A schematic picture of how scars can be used for state preparation is shown in Fig.~\ref{fig:Fig7}(D): an initial state is prepared and a fast diabatic evolution brings the state to the $\overline{\mathbb{Z}_2}$ phase by sweeping the detuning $\delta$. A global quench to a detuning $\delta$\,$=$\,$0$ then invokes collective nonergodic dynamics. 
While the position of the first revival occurs at time $T\sim\mathcal{O}(1)$, the size of the revival is expected to vanish in the thermodynamic limit. We consider the scaling of the decay and compare the data for the PXP chain and the $\sqrt{2}$-PXP chain in Fig.~\ref{fig:Fig7}(E). From the eigenvalue overlaps [Fig.~\ref{fig:Fig7}(A)] and a simple two-site dimer model (App.~\ref{app:dimer}), it is expected that signatures of scars are weaker in the $\sqrt{2}$-PXP chain. Indeed, we observe that the first revival peak vanishes with a faster exponential slope when the Rabi frequency is locally enhanced. Nevertheless, for the system sizes considered, the absolute fidelities are still remarkably large. Contrasted with the superexponential vanishing of the adiabatic gap, this opens up the possibility to cross into the $\mathbb{Z}_2$ phase by a quantum quench instead of a very slow adiabatic evolution.

As the perfect initial state for the quench might itself be difficult to prepare, we consider instead an intermediate state in the quasi-adiabatic evolution after the critical point (this stopping time would need to be fine-tuned). In the Appendix, we show such a realistic quench scenario after a quasi-adiabatic sweep for a chain of length $L=11$ with long-ranged tails and observe a significant state-transfer fidelity, albeit less pronounced than for a perfect initial $\overline{\mathbb{Z}_2}$ state. Here, the quench is performed after the quasi-adiabatic evolution. Related schemes where a quench is introduced in the middle of the quasi-adiabatic sweep can yield superior results in practical experimental setups~(cf.~Ref.~\cite{Lukin2022Exploring}). 
Generally, the readout is required to be timed to maximize the probability of observing the desired $\mathbb{Z}_2$ state. In this fashion, quantum quenches may be used advantageously as an algorithmic tool.

\subsection{Extension to quasi-2D grids} \label{ssec:2D}

\begin{figure}[t]
    \includegraphics[width=1.0\linewidth]{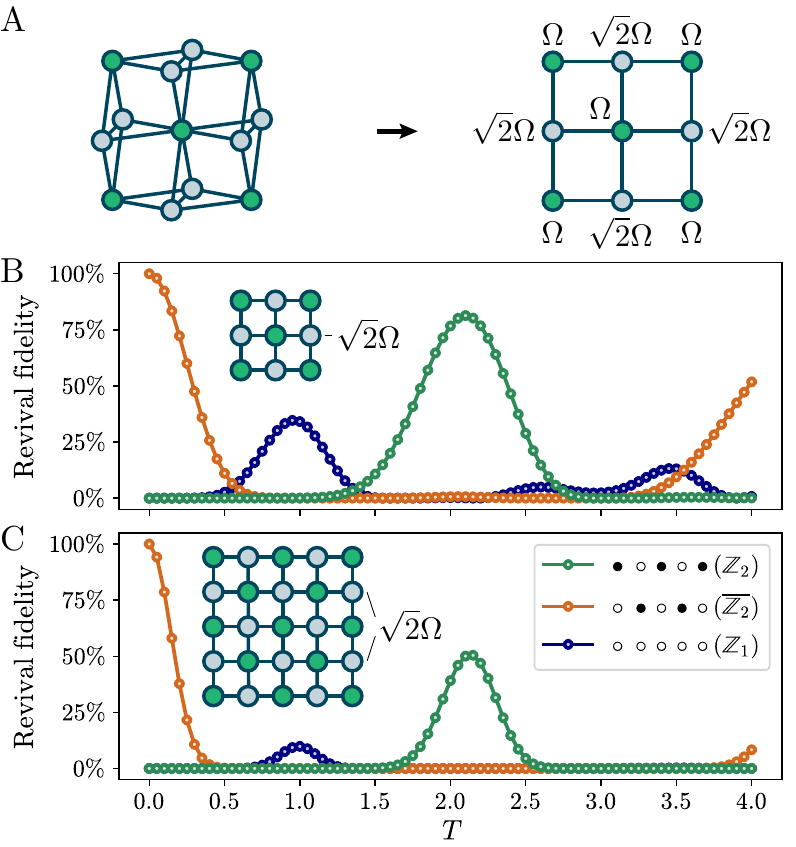}
    \caption{\textbf{(A)} A quasi-2D geometry which naturally extends the quasi-1D chain and maps to a 2D lattice with the Rabi frequency enhanced by a factor of $\sqrt{2}$ on every light blue site that is not a part of the nondegenerate MIS solution (green). \textbf{(B)} Revival fidelity for the imbalanced 3$\times$3 grid with an enhanced Rabi frequency on every light blue site. Initializing a quench with $\delta=0$ in the $\overline{\mathbb{Z}_2}$ state, we find a strong revival of the $\mathbb{Z}_2$ state, i.e., the MIS state, shortly after $T=2$. \textbf{(C)} Revival fidelity for the imbalanced 5$\times$5 grid (i.e., $N=37$ sites in the original quasi-2D grid), still exhibiting a significant peak of 50\%.}
    \label{fig:Fig8}
\end{figure}

The construction of the quasi-1D chain can be directly extended to a quasi-2D lattice. Given a square grid, the vertices that are not to be part of the MIS are replaced by doublets. An example constructed from a $3\times3$ grid is shown in Fig.~\ref{fig:Fig8}(A), featuring $N=13$ vertices. Importantly, our analysis of the spectral gap's scaling remains the same for this $\sqrt{2}$-PXP grid. For notational convenience, we continue to denote the single MIS configuration with $\mathbb{Z}_2$ (MIS with green nodes) and the complementary state by $\overline{\mathbb{Z}_2}$, corresponding to the light blue nodes. Even for this $\sqrt{2}$-PXP grid, the feature of quantum many-body scars persist, which we verify numerically. In our simulations, we simulate quenches to a detuning $\delta=0$ with doublet grids of $N=13$ ($3\times3$) and $N=37$ ($5\times5$) sites, shown in Fig.~\ref{fig:Fig8}(B,C). The quenches are initialized in the $\overline{\mathbb{Z}_2}$ state and show significant state-transfer fidelity to the $\mathbb{Z}_2$ state shortly after $T=2$, confirming that the doublet grid behaves similarly to the doublet chain.

\section{Discussion and outlook} \label{sec:discussion} \label{sec:Conclusion}

In this work, we introduced the concept of graphs with locally independent configurational choices and showed that it can cause a system to evolve into a configuration far from the global optimum in terms of Hamming distance at early times of an adiabatic evolution. 
This mechanism renders inefficient not only QAA but also simulated annealing (SA), an important classical multipurpose algorithm for solving optimization problems~\cite{Kirkpatrick1983Optimization}. In simulated annealing, the temperature of the thermal equilibrium state corresponding to the (classical) problem Hamiltonian is gradually lowered until the ground state is reached. In every iteration of this classical algorithm, a random spin (or cluster of spins) update is proposed and the update is accepted or rejected according to a rule that guarantees convergence to the Gibbs ensemble at the current temperature. In Ref.~\onlinecite{Ebadi2022Quantum}, it was observed that the ability of SA to find the solution of the MIS problem depends on the degeneracy ratio $R$: the expected hitting time for SA to prepare the MIS is bounded from below by a time proportional to $R$. For the chain of diamond plaquettes, $R$ increases exponentially with $N$ such that SA fails to solve the MIS problem on this graph efficiently. Hence, the local degrees of freedom make both QAA and SA ineffective for this model. 

We propose two approaches to mitigate the small adiabatic gap and to reduce the time to solution. Spin-exchange terms increase the adiabatic gap by orders of magnitude without changing the scaling of how the gap closes with system size. Moreover, using a Laplacian formalism, an exponential scaling of the spectral gap can be recovered. In the second approach, we considered quenches with the rapidly prepared suboptimal configuration as an initial state. Similar to the regular PXP chain, the $\sqrt{2}$-chain features signatures of quantum many-body scars, which result in oscillations from the suboptimal configuration to the target configuration (MIS). Even though the revival fidelities are expected to vanish in the thermodynamic limit, they remain significant for relatively large systems. We confirm an exponential scaling of this decay in the numerically accessible regime, suggesting a favorable asymptotic scaling compared to the superexponential scaling of the adiabatic gap.  We highlight that the nonergodic behavior of many-body scars may be stabilized such that the revival fidelity has a more favorable decay or does not decay at all in the thermodynamic limit. Concretely, it was shown for the PXP model that quasilocal deformations can make revivals virtually perfect~\cite{choi_emergent_2019}. Optimizing the initial state for the quench dynamics can potentially also improve the stability of the scars. This opens up possible intersections with optimal control theory to determine efficient protocols for state preparation on graph instances where the adiabatic gap becomes prohibitively small~\cite{caneva_optimal_2009}. 
It will be interesting to explore whether the combination of fast, nonadiabatic sweeps and quenches can give rise to a speedup over QAA more generally.

In our analysis, we start from the doublet Rydberg chain and show that in the limit of strong, short-ranged interactions, it can be described by a $\sqrt{2}$-Rydberg chain with a superexponentially small adiabatic gap. Besides this \emph{quantum} mechanism that relies on the coherent superposition of the doublets, there also exists a \emph{classical} mechanism that leads to a superexponentially small gap.
The classical mechanism is due to the tails of the long-range interaction, which favor a state in which the distance between Rydberg excitations is maximized by separating them diagonally across adjacent doublets. 
The resulting zigzag-type state also has a large Hamming distance from the MIS state. As the net difference in interaction energy between the zigzag-type state and the MIS is proportional to the chain length, the first-order transition happens later in the sweep for longer chains. Hence, the energy splitting due to the interaction tails induces a classical mechanism with a similar effect on the gap as the enhanced Rabi frequency in the $\sqrt{2}$-chain. For longer chains, the classical mechanism eventually dominates over the quantum mechanism. In the Appendix, we numerically confirm the predicted gap scaling in the doublet Rydberg model and compare it to the $\sqrt{2}$-Rydberg chain, where only the quantum mechanism is present.
The experimental implications of the interaction tails of the doublet Rydberg chain will be studied in future work~\cite{Lukin2022Exploring}, specifically, in the limit where the reduced chain approximation is no longer applicable. We note that by placing the atoms of the doublet sufficiently close, the physical properties of the $\sqrt{2}$-Rydberg chain, as investigated in this paper, are retained.

Our observations provide important insights relevant for practical realizations of optimization problems in Rydberg systems. 
Since local degrees of freedom arise in the cliques of a graph, it is likely that this mechanism will also impede the solution of the MIS problem on a more general set of graphs with many cliques. Further work is needed to ascertain to what extent local degeneracy and the vdW tails are responsible for the observed correlation between the degeneracy ratio and the adiabatic gap in random graphs~\cite{Ebadi2022Quantum}. Our work opens the door to further theoretical and experimental studies of the characteristics that render instances of the MIS problem hard to solve with QAA. An improved understanding of the connection between graph geometries and the adiabatic gap will guide the search for classes of instances for which optimization with Rydberg atom arrays may achieve a practical quantum speedup.\\

\begin{acknowledgments}
The authors thank J.~Ignacio Cirac, Jin-Guo Liu, Alexander Lukin, and Sheng-Tao Wang for insightful discussions.
B.F.S.~acknowledges funding from the European Union’s Horizon 2020 research and innovation program under Grant No.~899354 (FET Open SuperQuLAN). D.S.W.~acknowledges funding from the European Union's Horizon 2020 research and innovation programme under the Marie Skłodowska-Curie Grant No.~101023276. This research is part of the Munich Quantum Valley, which is supported by the Bavarian state government with funds from the Hightech Agenda Bayern Plus. We further acknowledge the US Department of Energy (DOE Quantum Systems Accelerator Center, contract number 7568717 and DE-SC0021013), the Center for Ultracold Atoms, the National Science 
Foundation, the Army Research Office MURI (grant number W911NF-20-1-0082),
and the DARPA ONISQ program (grant number W911NF2010021). N.M.~acknowledges support from the Department of Energy Computational Science Graduate Fellowship under Award Number DE-SC0021110. M.C.~acknowledges support from Department of Energy Computational Science Graduate Fellowship under Award Number (DESC0020347).  R.S.~is supported by the Princeton Quantum Initiative Fellowship.
\end{acknowledgments}

\bibliography{biblio}
\clearpage
\appendix
\section{Scaling of the adiabatic gap}

\subsection{Perturbation theory}
In this section, we seek to quantify the size of the smallest spectral gap along the adiabatic sweep to prepare the ground state of the $k$-PXP chain with the Rabi frequency enhanced by a factor $k$ on every even site. We denote $L$\,$=$\,$2K$\,$-$\,$1$, such that $K=|\text{MIS}|$, and make use of a perturbative approach in the limit of $\delta \gg \Omega$ and $\delta > 0$:
\begin{align}
    \frac{H_\text{eff}}{\delta} = H^{}_0 + \frac{\Omega}{\delta} H^{}_x = H^{}_0 + \epsilon H^{}_x.
\end{align}
For the doublet PXP chain, we have $k$\,$=$\,$\sqrt{2}$. The unperturbed Hamiltonian and the perturbation are given by
\begin{align}
    H^{}_0 &= - \sum_i^{L} n^{}_i, \quad \text{and}\\
    H^{}_x &= \sum_{i \; \text{odd}}^{L} P^{}_{i-1}\sigma_i^xP^{}_{i+1} + k \sum_{i \; \text{even}}^{L} P^{}_{i-1}\sigma_i^x P^{}_{i+1}
\end{align}
with $P_0$\,$=$\,$P_{L+1}$\,$=$\,$I$ at the boundaries. The eigenstates of the unperturbed Hamiltonian $H_0$ are classical configurations in the occupation basis. In Rayleigh-Schrödinger perturbation theory, the perturbed energies are given by
\begin{align}
    E^{}_n(\epsilon) = E_n^{(0)}  + \epsilon^2 \sum_{k\neq n} \frac{|\langle k^{(0)}|H_x|n^{(0)} \rangle|^2}{E_n^{(0)} - E_k^{(0)}} + \mathcal{O}(\epsilon^3)
\end{align}
because, at first order, all contributions vanish due to the orthogonality of the eigenstates. Hence, the perturbed energy for the nondegenerate ground state can be easily computed as $E_0 = -K + \epsilon^2 (-K)$. This is because the ground state of $H_0$ is the MIS with $K$ sites in the Rydberg state and there are $K$ terms in the sum contributing to the correction at second order: each of the $K$ occupied sites may become unoccupied individually through $H_x$, and there are no other classical configurations that the ground state connects to at this order.

The first excited state, however, is highly degenerate. Thus, we consider the matrix
\begin{align}
    M^{}_{ij} = -\epsilon^2 \sum_{l \in \{\text{MIS}, \text{MIS}-2\}} \frac{\langle i|H_x|l\rangle\langle l|H_x|j\rangle }{E_{\text{MIS}-1}^{(0)} - E_{l}^{(0)}}
\end{align}
where $i,j$\,$\in$\,$\{\text{MIS}$\,$-$\,$1\}$. We observe that the sorted diagonal of $M$ is given by
\begin{align}
    \text{diag}(M) = \epsilon^2\{\dots,k^2(K-2)+1,k^2(K-1)\},
\end{align}
where the last element corresponds to the energy of the MIS-1 state with the largest Hamming distance to the ground state, i.e., the $\overline{\mathbb{Z}_2}$ state. The penultimate element is one of the classical states with $K-2$ occupied even sites and one occupied odd site. The MIS-1 state $\overline{\mathbb{Z}_2}$ couples to only two other MIS-1 states through edge excitations: as the state is a maximal independent set state, an occupied spin needs to be moved from the second to the first or from the second-last to the last site of the chain. Hence, the matrix $M$ has only two off-diagonal elements in the last row (or column) which have value $k$. By the same argument, we have three elements of value $k$ in the second-last row.\\
Clearly, if $K\rightarrow\infty$, the largest eigenvalue of $M$ is given by $k^2(K-1)$. For sufficiently large $k$ as well, we can argue why the eigenvalue would indeed be close to this value. To do so, we make use of the Gershgorin circle theorem~\cite{gershgorin1931uber}.
\begin{theorem}[Gershgorin circle]
Every eigenvalue of a square matrix $A$ with entries $a_{ij}$ lies within at least one of the Gershgorin discs $D(a_{ii}, R_i)$ centered at $a_{ii}$ with radius $R_i=\sum_{j\neq i}|a_{ij}|$. If a single disc is disjoint from the other discs, it contains exactly one eigenvalue of $A$.
\end{theorem}
The discs corresponding to the largest two eigenvalues are $D(k^2(K-1),2k)$ and $D(k^2(K-2)+1,3k)$, respectively. Hence, these discs are disjoint for $\tilde k>(\sqrt{29}+5)/2\approx 5.19$. This implies that the proof here does not yet apply for $k=\sqrt{2}$; however, numerical evidence (see Fig.~\ref{fig:sizeposition}) suggests a similar behavior can indeed be expected for values of $k$ that are significantly larger than 1 but smaller than $\tilde k$.\\
Having established that the largest eigenvalue $r$ concentrates around $r\approx k^2(K-1)$ for sufficiently large $k$, independent of the chain length, we analyze the dominant eigenvector $v$ using the Perron-Frobenius theorem~\cite{meyer2000matrix}.
\begin{theorem}[Perron-Frobenius]
For a square, positive, symmetric matrix $A$ with dominant eigenvalue $r$ and normalized dominant eigenvector $v$, the relation $\lim_{m\rightarrow \infty} A^m/r^m = vv^T$ holds.
\end{theorem}
Thus, the Perron-Frobenius theorem implies that the dominant eigenvector approaches $v$\,$=$\,$(0,\dots,0,1)$ as $K$ or $k$\,$\rightarrow$\,$\infty$. The configuration corresponding to the dominant eigenvector is the MIS-1 state with maximal Hamming distance ($\overline{\mathbb{Z}_2}$) from the MIS state ($\mathbb{Z}_2$). In other words, the MIS-1 state is localized in configuration space at the critical point.\\
We consider the tunneling matrix element between the two localized states to compute the adiabatic gap. As they couple only via $L^\mathrm{th}$-order quantum fluctuations, the spectral gap is given as 
\begin{align}
\Delta^{}_\text{min} &= \epsilon^{L}_\text{crit} \sum_{\boldsymbol{\lambda}} \prod_{\lambda_i} \sqrt{2}^{1+(-1)^i} \frac{1}{E_{\lambda_i}} \\
&= \epsilon^{2K-1}_\text{crit} \sqrt{2}^{K-1} \sum_{\boldsymbol{\lambda}} \prod_{\lambda_i} \frac{1}{E_{\lambda_i}},
\end{align}
where $E_{\lambda_i}$ is the energy denominator from the perturbative expansion and the sum runs over all trajectories $\boldsymbol{\lambda}$ where $L$ sites are flipped from unoccupied to occupied (and vice versa) to connect the $\mathbb{Z}_2$ state to the $\overline{\mathbb{Z}_2}$ state. The position of the minimum spectral gap $\epsilon_\text{crit}$ is computed using a mean-field ansatz (see Sec.~\ref{sec:MF}). We can also confirm the location of the critical point using a perturbative argument where we consider the limit of large $K$ and compute the crossing between the energies of the MIS state and the MIS-1 state that is closest in energy to obtain $\epsilon_\text{crit}$\,$=$\,$1/\sqrt{k^2(K-1)-K}$. The perturbative argument holds as the critical point occurs in the classical regime for large system sizes or large $k$. The gap will be given by $\Delta_\text{min}$\,$=$\,$w(K) (\sqrt{k^2(K-1)-K})^{-L}$, where $w(K)$ is a correction factor. As in Ref.~\onlinecite{Altshuler_2010}, we have a total number of trajectories which goes as $\mathcal{O}(K!)$ that cancels the contribution from the $E_{\lambda_i}$, which also gives  $\mathcal{O}(K!)$, such that $w(K)$ does not change the scaling of the gap. Hence, we find that the adiabatic gap vanishes as 
$ \Delta_\text{min} \sim \mathcal{O}(\exp(-L\log(L))$.

\subsection{Validity of the reduced chain approximation} \label{app:ssec:validity}

\begin{figure}[b]
    \includegraphics[width=1.0\linewidth]{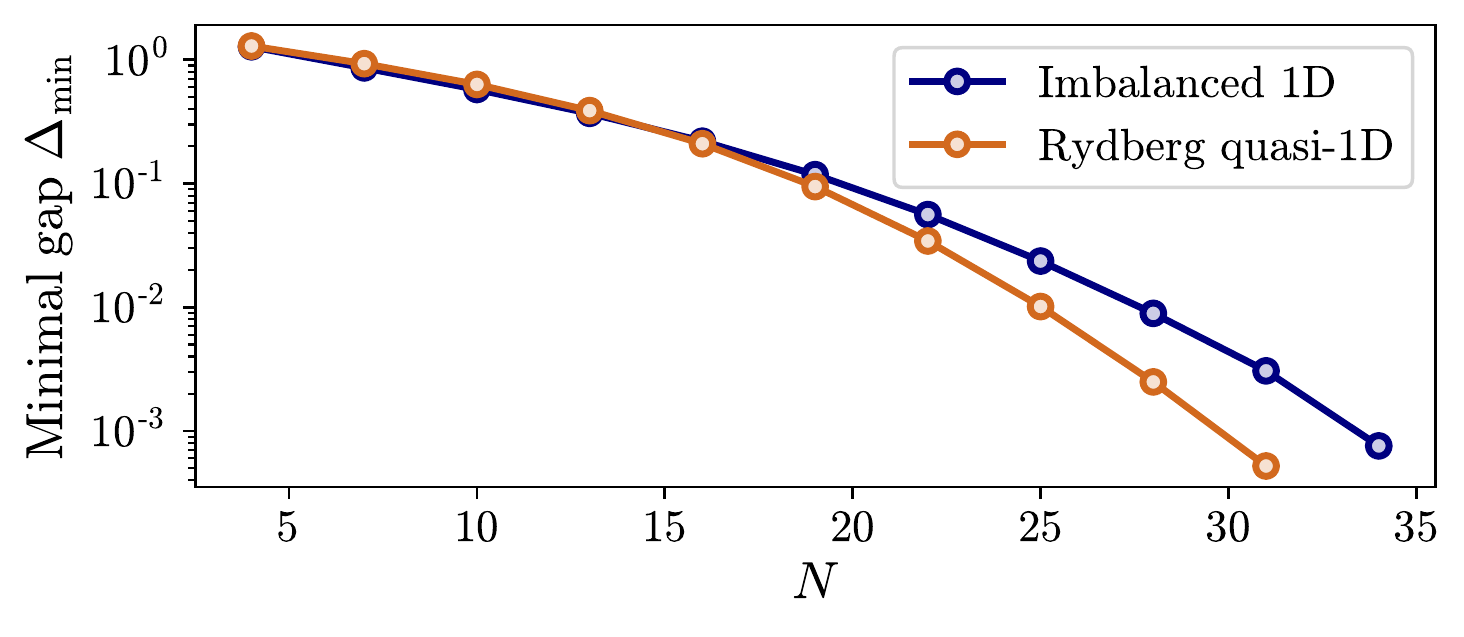}
    \caption{Superexponentially decreasing minimum gap for the doublet Rydberg model in the geometry described in Sec.~\ref{app:ssec:validity} retaining interaction tails up to distances $<2.5$ (orange), and the $\sqrt{2}$-Rydberg model with $k$\,$=$\,$\sqrt{2}$ but no tails (blue); $V$ is chosen to be $100$ for both cases. The steeper slope for the doublet Rydberg model is attributed to the vdW tails favoring MIS-1 configurations with large Hamming distance.
    }
    \label{fig:CompareRyd}
\end{figure}

We consider the spectral gap of the doublet Rydberg model (including next-nearest neighbors separated by distances $<2.5$) and compare it to that of the $\sqrt{2}$-Rydberg chain with an imbalanced Rabi frequency (but without long-ranged interaction tails) to validate that the approximation of the reduced chain is suitable for the geometries considered. The specific geometry considered here has doublet sites spaced by $1$ and the single atom at distance $\sqrt{1.5}$ from each doublet atom. The interaction coefficient is chosen to be $V$\,$=$\,$100$ for both chains. Indeed, we see that the scaling of the minimum gap is superexponential in both cases in Fig.~\ref{fig:CompareRyd}, although there are quantitative differences for longer chains.   Here, the blue line is reproduced from Fig.~\ref{fig:Fig5}. The difference can be explained as follows: due to the tails, a distinct MIS-1 configuration emerges which is the ground state within a range of positive $\delta/\Omega$ before the critical detuning. In this case, an excitation on one doublet is perfectly correlated with another in a diagonally adjacent position, giving rise to a superposition of two zigzag patterns of Rydberg excitations. The reduced chain approximation is valid when the two atoms of the doublet are placed sufficiently close to each other. The gap closing due to a classical energy splitting is analyzed in detail in~\cite{Lukin2022Exploring}.

\section{Variable imbalance parameter $k$}

\begin{figure}[tb]
    \includegraphics[width=1.0\linewidth]{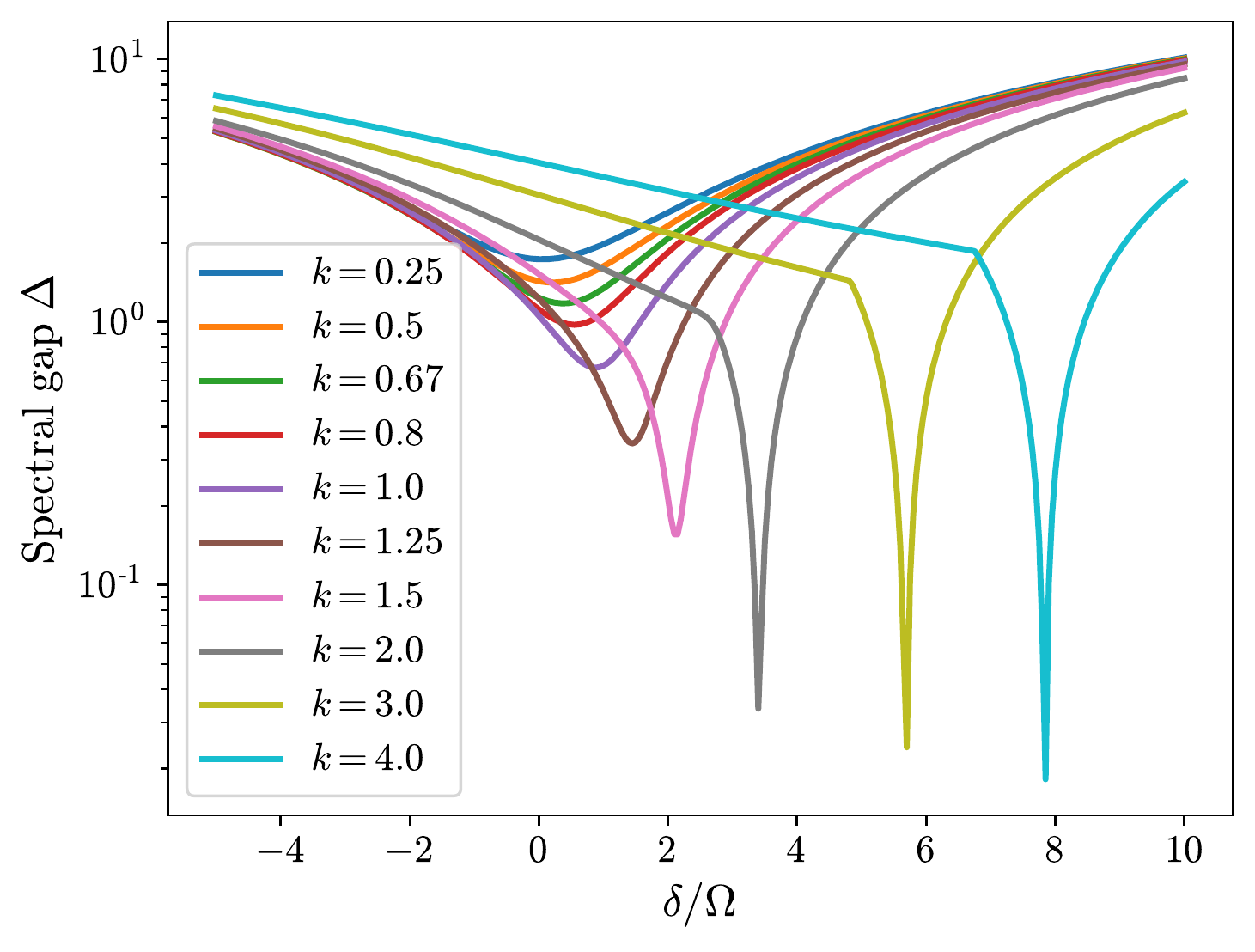}
    \caption{Spectral gap for different values of $k$ for a chain of length $L=11$. The Rabi frequency is set to $\Omega=1$ and is enhanced by the factor $k$ on every even site. The critical detuning is found to be at larger values of $\delta$ when increasing $k$. 
    }
    \label{fig:Gapvaryk}
\end{figure}

It is insightful to consider the closing of the spectral gap for different enhancement factors $k$ as a function of $\delta/\Omega$. We plot the gap for a chain of length $L=11$ in Fig.~\ref{fig:Gapvaryk} and observe that for increasing $k$, the position of the minimum shifts toward larger $\delta$ and, importantly, the gap itself closes more sharply. For $k=1$, we know that the phase transition is of second order, while it becomes a first-order transition for sufficiently large $k$.

\label{app:gadget}

\begin{figure}[b]
    \includegraphics[width=1.0\linewidth]{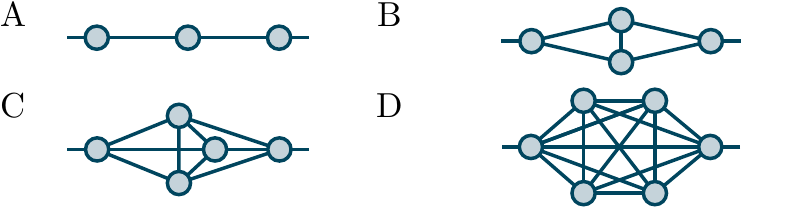}
    \caption{Variants of a gadget introduced into a 1D chain in \textbf{(A)}. The doublet nodes in \textbf{(B)} lead to a local enhancement of the Rabi frequency of $k$\,$=$\,$\sqrt{2}$, but other gadgets can be imagined, e.g., inserting a fully-connected graph of three vertices in \textbf{(C)} for $k$\,$=$\,$\sqrt{3}$, or four vertices in \textbf{(D)} for $k$\,$=$\,$2$.}
    \label{fig:gadget}
\end{figure}

In the main text, we mention that while it might not be feasible to tune $k$ continuously, it is possible to implement different values of $k$ by using gadgets~\cite{Nguyen2022Quantum}, which are graphs with certain properties replacing a single site (or multiple sites) in the original graph. In Fig.~\ref{fig:gadget}, we present examples of how the quasi-1D chain can be extended for higher $k$: instead of a two-vertex clique, a clique of three or four vertices might be inserted, yielding $k=\sqrt{3}$ and $k=\sqrt{4}=2$, respectively. All vertices from each clique connect to the previous and subsequent vertex of the chain.
For the 3-vertex clique [panel (C)], the symmetric subspace spans $\{\ket{ggg},(\ket{ggr} +\ket{grg}+\ket{rgg})/\sqrt{3}\}$, thus leading to the enhancement factor of $k=\sqrt{3}$.

\section{Quantum many-body scars in the $\sqrt{2}$-PXP chain}

\subsection{Approximate revival time in the dimer model}\label{app:dimer}

We would like to estimate the time that maximizes the fidelity of the $\mathbb{Z}_2$ state on quenching from the $\overline{\mathbb{Z}_2}$ state in the PXP chain where every even site has an enhanced Rabi frequency $k\Omega$. Here, we consider a simple model of two-site dimers and extend the ansatz from Ref.~\onlinecite{Bernien2017Probing} to the $\sqrt{2}$-PXP chain. The Hamiltonian considered is 
\begin{align}
    H^{}_q = \Omega\sum_{i\ \text{odd}}^L P^{}_{i-1}\sigma_x^iP^{}_{i+1} +  k\Omega\sum_{i\ \text{even}}^L P^{}_{i-1}\sigma_x^iP^{}_{i+1}
\end{align}
where, as before, $P_0$\,$=$\,$P_{L+1}$\,$=$\,$I$. We consider an ansatz $\ket{\psi (t)}= \otimes_{1\le j \le L/2} \ket{\phi(t)}_{2j-1, 2j}$ describing independent two-site dimers. Within each dimer, the blockade constraint is enforced, thus allowing only superpositions of the basis states $\ket{rg}$, $\ket{gr}$ and $\ket{gg}$. Note that the blockade between adjacent dimers, however, is ignored here; hence, this model can only be a first-order approximation to the true quantum dynamics. We find that for an initial state $\ket{\phi(t=0)} = \ket{gr}$, the time evolved state can be described by
\begin{align}
\nonumber
    \ket{\phi(t)} &= \frac{ik \sin(\sqrt{1+k^2}t\Omega)}{\sqrt{1+k^2}} \ket{gg} \\
    \nonumber
    &+ \frac{1+k^2 \cos(\sqrt{1+k^2}t\Omega)}{1+k^2} \ket{gr} \\
    &+ \frac{-k+k \cos(\sqrt{1+k^2}t\Omega)}{1+k^2} \ket{rg}.
\end{align}
Hence, the revival time to quench into the state $\ket{rg}$ is  
\begin{align}
\tilde{T}^{}_k=\frac{2}{\sqrt{1+k^2}}\frac{\pi}{2\Omega}.
\end{align}
This simple model shows that the periodic evolution of the revivals is faster for larger imbalances. It also predicts that the fidelity of the \textit{state transfer} from $\ket{gr}$ to $\ket{rg}$ is below one for $k>1$, contributing to a decaying signal for larger system sizes.

\subsection{Decay of the revival fidelity and simulation of quench dynamics with vdW tails.}

\begin{figure}[h]
    \includegraphics[width=1.0\linewidth]{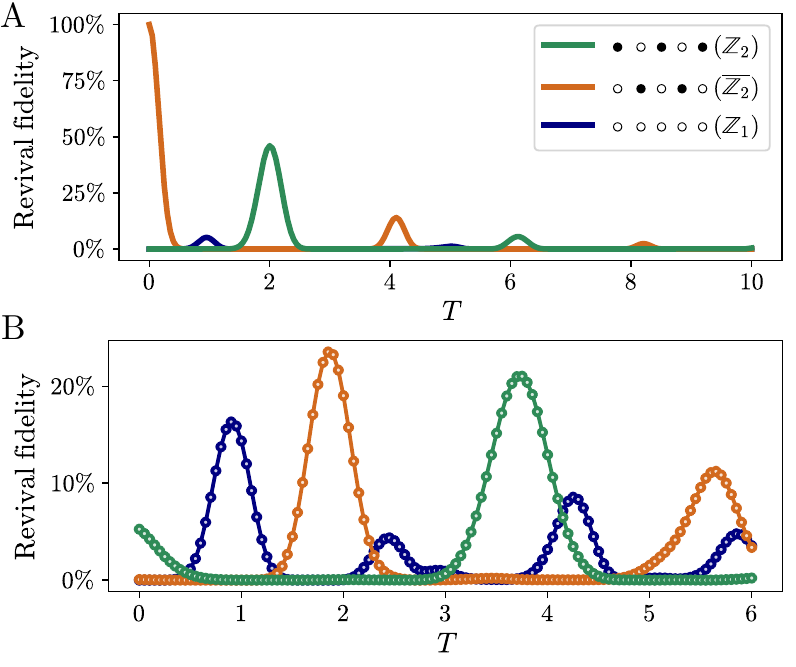}
    \caption{\textbf{(A)} Revival and state-transfer fidelity for the $\sqrt{2}$-PXP chain as in Fig.~\ref{fig:Fig8}, but for longer quench times. \textbf{(B)} For the $\sqrt{2}$-PXP chain ($L=11$), the revivals are shown after a linear quasi-adiabatic sweep with $\delta\in[-40,40]$, $\Omega$=1, total evolution time $T_\text{QAA}=50$, and tails (corresponding to $V\,=\,20$), aborted at 73\% of the sweep; the peak $\mathbb{Z}_2$ fidelity is 21\%.}
    \label{fig:Long_SQS}
\end{figure}

In order to quench from the $\overline{\mathbb{Z}_2}$ state to the $\mathbb{Z}_2$ state, we are mostly interested in the fidelity of the first revival peak. For longer quench times, more revivals are observed; however, they decay rather quickly in the models considered. For completeness, we plot in Fig.~\ref{fig:Long_SQS}(A) the evolution for longer times where the decay of the revivals can be observed.

We propose that quench dynamics can be used as an algorithmic tool for state preparation. An example scenario where a quench follows a quasi-adiabatic evolution is simulated in Fig.~\ref{fig:Long_SQS}(B). Here, to study the dynamics, we consider the $\sqrt{2}$-PXP chain and add vdW interaction tails corresponding to an interaction strength $V\,=\,20$. As mentioned in the main text, related schemes where the quench is introduced in the middle of a quasi-adiabatic can yield superior results and are investigated in Ref.~\cite{Lukin2022Exploring}. 

\end{document}